\documentclass[acus]{pac99}
\usepackage{epsfig}
\newcommand{\ud}{\mathrm{d}}
\begin{document}
\title{NONLINEAR ACCELERATOR PROBLEMS VIA WAVELETS:\\
6. REPRESENTATIONS AND QUASICLASSICS VIA FWT}
\author{A.~Fedorova,  M.~Zeitlin, IPME, RAS, St.~Petersburg, Russia
  \thanks{ e-mail: zeitlin@math.ipme.ru}
  \thanks{http://www.ipme.ru/zeitlin.html:
          http://www.ipme.nw.ru/zeitlin.html}}
\maketitle
\begin{abstract}
In this series of eight papers  we
present the applications of methods from
wavelet analysis to polynomial approximations for
a number of accelerator physics problems.
In this part we consider application of FWT to
metaplectic representation(quantum and chaotical problems)
and quasiclassics.
\end{abstract}

\section{INTRODUCTION}
This is the sixth part of our eight presentations in which we consider
applications of methods from wavelet analysis to nonlinear accelerator
physics problems.
 This is a continuation of our results from [1]-[8],
in which we considered the applications of a number of analytical methods from
nonlinear (local) Fourier analysis, or wavelet analysis, to nonlinear
accelerator physics problems
 both general and with additional structures (Hamiltonian, symplectic
or quasicomplex), chaotic, quasiclassical, quantum. Wavelet analysis is
a relatively novel set of mathematical methods, which gives us a possibility
to work with well-localized bases in functional spaces and with the
general type of operators (differential, integral, pseudodifferential) in
such bases.
In contrast  with parts 1--4 in parts 5--8 we try to take into account
before using power analytical approaches underlying algebraical, geometrical,
topological structures related to kinematical, dynamical and hidden
symmetry of physical problems.
 In part 2
according to the orbit method and by using construction
from the geometric quantization theory we construct the symplectic
and Poisson structures associated with generalized wavelets by
using metaplectic structure.
In part 3  we consider applications of very
useful fast wavelet transform technique (FWT) (part 4) to calculations in
quasiclassical evolution dynamics.
This method gives maximally sparse representation of (differential) operator
 that allows us to take into account contribution from each level of
resolution.

\section{Metaplectic Group and Representations}
Let $Sp(n)$ be
symplectic group, $Mp(n)$ be its unique two- fold covering --
metaplectic group [9].   Let V be a symplectic vector space
with symplectic form ( , ), then $R\oplus V$ is nilpotent Lie
algebra - Heisenberg algebra:
$[R,V]=0, \ [v,w]=(v,w)\in
R,\  [V,V]=R$.
$Sp(V)$ is a group of automorphisms of
Heisenberg algebra.
Let N be a group with Lie algebra $R\oplus
V$, i.e.  Heisenberg group.  By Stone-- von Neumann theorem
Heisenberg group has unique irreducible unitary representation
in which $1\mapsto i$. Let us also consider the  projective
representation of symplectic group $Sp(V)$:
$U_{g_1}U_{g_2}=c(g_1,g_2)\cdot U_{g_1g_2}$, where c is a map:
$Sp(V)\times Sp(V)\rightarrow S^1$, i.e. c  is $S^1$-cocycle.
But this representation is unitary representation of universal
covering, i.e. metaplectic group $Mp(V)$. We give this
representation without Stone-von Neumann theorem.\
Consider a new group $F=N'\bowtie Mp(V),\quad \bowtie$ is semidirect
product (we consider instead of  $ N=R\oplus V$ the $
N'=S^1\times V, \quad S^1=(R/2\pi Z)$). Let $V^*$ be dual to V,
$G(V^*)$ be automorphism group of $V^*$.Then F is subgroup of $
G(V^*)$, which consists of elements, which acts on $V^*$ by affine
transformations.
This is the key point!
Let $q_1,...,q_n;p_1,...,p_n$ be symplectic basis in V,
$\alpha=pdq=\sum p_{i}dq_i $  and $d\alpha$ be symplectic form on
$V^*$. Let M be fixed affine polarization, then for $a\in F$ the
map $a\mapsto \Theta_a$ gives unitary representation of G:
$ \Theta_a: H(M) \rightarrow H(M)$.
Explicitly  we have for representation of N on H(M):
$
(\Theta_qf)^*(x)=e^{-iqx}f(x),  \
 \Theta_{p}f(x)=f(x-p)
$
The representation of N on H(M) is irreducible. Let $A_q,A_p$
be infinitesimal operators of this representation
$$
 A_q=\lim_{t\rightarrow 0} \frac{1}{t}[\Theta_{-tq}-I], \quad
 A_p=\lim_{t\rightarrow 0} \frac{1}{t}[\Theta_{-tp}-I],
$$
$$\mbox{then }
A_q f(x)=i(qx)f(x),\  A_p f(x)=\sum p_j\frac{\partial
f}{\partial x_j}(x)
$$
Now we give the representation of infinitesimal ba\-sic
elements. Lie algebra of the group F is the algebra of all
(non\-ho\-mo\-ge\-ne\-ous) quadratic po\-ly\-no\-mi\-als of (p,q) relatively
Poisson bracket (PB). The basis of this algebra consists of
elements
$1,q_1,...,q_n$,\  $p_1,...,p_n$,\ $ q_i q_j, q_i p_j$,\ $p_i p_j,
 \quad i,j=1,...,n,\quad i\leq j$,
 \begin{eqnarray*}
 & &PB \ is
\quad \{ f,g\}=\sum\frac{\partial f}{\partial p_j}
\frac{\partial g}{\partial q_i}-\frac{\partial f}{\partial q_i}
\frac{\partial g}{\partial p_i} \\
 & &\mbox{and}  \quad
   \{1,g \}= 0 \quad for \mbox{ all} \ g,\\
& &\{ p_i,q_j\}= \delta_{ij},\quad \{p_i
q_j,q_k\}=\delta_{ik}q_j,\\
 & &   \{p_i q_j,p_k\}=-\delta_{jk}p_i, \quad \{p_ip_j,p_k\}=0,\\
& & \{p_i p_j,q_k \}=
\delta_{ik}p_j+\delta_{jk}p_i,\quad
  \{ q_i q_j,q_k\}=0,\\
& & \{q_i q_j,p_k\}=-\delta_{ik}q_j-\delta_{jk}q_i
 \end{eqnarray*}
so, we have the representation of basic elements
 $ f\mapsto A_f : 1\mapsto i, q_k\mapsto ix_k $,
\begin{eqnarray*}
&& p_l\mapsto\frac{\partial}{\partial x^l}, p_i q_j\mapsto
x^i\frac{\partial}{\partial x^j}+\frac{1}{2}\delta_{ij},\\
&& p_k p_l\mapsto \frac{1}{i}\frac{\partial^k}{\partial x^k\partial
x^l}, q_k q_l\mapsto ix^k x^l
\end{eqnarray*}
This gives  the structure of the Poisson mani\-folds to
representation of any (nilpotent) algebra or in other words to
continuous wavelet trans\-form.
According to this approach we
can construct by using methods of geometric quantization theory
many "symplectic wavelet constructions" with corresponding
symplectic or Poisson structure on it.
Then we may produce symplectic invariant wavelet calculations for
PB or commutators which we may use in quantization procedure or
in chaotic dynamics (part 8) via operator representation from
section 4.

\section{Quasiclassical Evolution}

Let us consider classical and quantum dynamics in phase space
$\Omega=R^{2m}$ with coordinates $(x,\xi)$ and generated by
Hamiltonian ${\cal H}(x,\xi)\in C^\infty(\Omega;R)$.
If $\Phi^{\cal H}_t:\Omega\longrightarrow\Omega$ is (classical) flow then
time evolution of any bounded classical observable or
symbol $b(x,\xi)\in C^\infty(\Omega,R)$ is given by $b_t(x,\xi)=
b(\Phi^{\cal H}_t(x,\xi))$. Let $H=Op^W({\cal H})$ and $B=Op^W(b)$ are
the self-adjoint operators or quantum observables in $L^2(R^n)$,
representing the Weyl quantization of the symbols ${\cal H}, b$ [9]
\begin{eqnarray*}
&&(Bu)(x)=\frac{1}{(2\pi\hbar)^n}\int_{R^{2n}}b\left(\frac{x+y}{2},\xi\right)
\cdot\\
&&e^{i<(x-y),\xi>/\hbar}u(y)\ud y\ud\xi,
\end{eqnarray*}
where $u\in S(R^n)$ and $B_t=e^{iHt/\hbar}Be^{-iHt/\hbar}$ be the
Heisenberg observable or quantum evolution of the observable $B$
under unitary group generated by $H$. $B_t$ solves the Heisenberg equation of
motion
$\dot{B}_t=({i}/{\hbar})[H,B_t].$
Let $b_t(x,\xi;\hbar)$ is a symbol of $B_t$ then we have
 the following equation for it
\begin{equation}
\dot{b}_t=\{ {\cal H}, b_t\}_M,
\end{equation}
with the initial condition $b_0(x,\xi,\hbar)=b(x,\xi)$.
Here $\{f,g\}_M(x,\xi)$ is the Moyal brackets of the observables
$f,g\in C^\infty(R^{2n})$, $\{f,g\}_M(x,\xi)=f\sharp g-g\sharp f$,
where $f\sharp g$ is the symbol of the operator product and is presented
by the composition of the symbols $f,g$
\begin{eqnarray*}
&&(f\sharp g)(x,\xi)=\frac{1}{(2\pi\hbar)^{n/2}}\int_{R^{4n}}
e^{-i<r,\rho>/\hbar+i<\omega,\tau>/\hbar}\\
&& \cdot f(x+\omega,\rho+\xi)\cdot
g(x+r,\tau+\xi)\ud\rho \ud\tau \ud r\ud\omega.
\end{eqnarray*}
For our problems it is useful that $\{f,g\}_M$ admits the formal
expansion in powers of $\hbar$:
\begin{eqnarray*}
&&\{f,g\}_M(x,\xi)\sim \{f,g\}+2^{-j}\cdot\\
&&\sum_{|\alpha+\beta|=j\geq 1}(-1)^{|\beta|}\cdot
(\partial^\alpha_\xi fD^\beta_x g)\cdot(\partial^\beta_\xi
 gD^\alpha_x f),
\end{eqnarray*}
 where $\alpha=(\alpha_1,\dots,\alpha_n)$ is
a multi-index, $|\alpha|=\alpha_1+\dots+\alpha_n$,
$D_x=-i\hbar\partial_x$.
So, evolution (1) for symbol $b_t(x,\xi;\hbar)$ is
\begin{eqnarray}
&&\dot{b}_t=\{{\cal H},b_t\}+\frac{1}{2^j}
\sum_{|\alpha|+\beta|=j\geq 1}(-1)^{|\beta|}
\cdot\\
&&\hbar^j
(\partial^\alpha_\xi{\cal H}D_x^\beta b_t)\cdot
(\partial^\beta_\xi b_t D_x^\alpha{\cal H}).\nonumber
\end{eqnarray}

At $\hbar=0$ this equation transforms to classical Liouville equation
\begin{equation}
\dot{b}_t=\{{\cal H}, b_t\}.
\end{equation}
Equation (2) plays a key role in many quantum (semiclassical) problem.
We note only the problem of relation between quantum and classical evolutions
or how long the evolution of the quantum observables is determined by the
corresponding classical one [9].
Our approach to solution of systems (2), (3) is based on our technique
from [1]-[8] and very useful linear parametrization for differential operators
which we present in the next section.

\section{FAST WAVELET TRANSFORM FOR DIF\-FE\-RENTIAL OPERATORS}

Let us consider multiresolution representation
$\dots\subset V_2\subset V_1\subset V_0\subset V_{-1}
\subset V_{-2}\dots$ (see our other papers from this series for
details of wavelet machinery). Let T be an operator $T:L^2(R)
\rightarrow L^2(R)$, with the kernel $K(x,y)$ and
$P_j: L^2(R)\rightarrow V_j$ $(j\in Z)$ is projection
operators on the subspace $V_j$ corresponding to j level of resolution:
$(P_jf)(x)=\sum_k<f,\varphi_{j,k}>\varphi_{j,k}(x).$ Let
$Q_j=P_{j-1}-P_j$ is the projection operator on the subspace $W_j$ then
we have the following "microscopic or telescopic"
representation of operator T which takes into account contributions from
each level of resolution from different scales starting with
coarsest and ending to finest scales [10]:
$
T=\sum_{j\in Z}(Q_jTQ_j+Q_jTP_j+P_jTQ_j).
$
We remember that this is a result of presence of affine group inside this
construction.
The non-standard form of operator representation [10] is a representation of
an operator T as  a chain of triples
$T=\{A_j,B_j,\Gamma_j\}_{j\in Z}$, acting on the subspaces $V_j$ and
$W_j$:
$
 A_j: W_j\rightarrow W_j, B_j: V_j\rightarrow W_j,
\Gamma_j: W_j\rightarrow V_j,
$
where operators $\{A_j,B_j,\Gamma_j\}_{j\in Z}$ are defined
as
$A_j=Q_jTQ_j, \quad B_j=Q_jTP_j, \quad\Gamma_j=P_jTQ_j.$
The operator $T$ admits a recursive definition via
$$T_j=
\left(\begin{array}{cc}
A_{j+1} & B_{j+1}\\
\Gamma_{j+1} & T_{j+1}
\end{array}\right),$$
where $T_j=P_jTP_j$ and $T_j$ works on $ V_j: V_j\rightarrow V_j$.
It should be noted that operator $A_j$ describes interaction on the
scale $j$ independently from other scales, operators $B_j,\Gamma_j$
describe interaction between the scale j and all coarser scales,
the operator $T_j$ is an "averaged" version of $T_{j-1}$.
The operators $A_j,B_j,\Gamma_j,T_j$ are represented by matrices
$\alpha^j, \beta^j, \gamma^j, s^j$
\begin{eqnarray}
\alpha^j_{k,k'}&=&\int\int K(x,y)\psi_{j,k}(x)\psi_{j,k'}(y)\ud x\ud y\nonumber\\
\beta^j_{k,k'}&=&\int\int K(x,y)\psi_{j,k}(x)\varphi_{j,k'}(y)\ud x\ud y\\
\gamma^j_{k,k'}&=&\int\int K(x,y)\varphi_{j,k}(x)\psi_{j,k'}(y)\ud x\ud y\nonumber\\
s^j_{k,k'}&=&\int\int K(x,y)\varphi_{j,k}(x)\varphi_{j,k'}(y)\ud x\ud y\nonumber
\end{eqnarray}
We may compute the non-standard representations of operator $\ud/\ud x$ in the
wavelet bases by solving a small system of linear algebraical
equations. So, we have for objects (4)
\begin{eqnarray*}
\alpha^j_{i,\ell}&=&2^{-j}\int\psi(2^{-j}x-i)\psi'(2^{-j}-\ell)2^{-j}\ud x\\
&=&2^{-j}\alpha_{i-\ell}\\
\beta^j_{i,\ell}&=&2^{-j}\int\psi(2^{-j}x-i)\varphi'(2^{-j}x-\ell)2^{-j}\ud x\\
&=&2^{-j}\beta_{i-\ell}\\
\gamma^j_{i,\ell}&=&2^{-j}\int\varphi(2^{-j}x-i)\psi'(2^{-j}x-\ell)2^{-j}\ud
x\\
&=&2^{-j}\gamma_{i-\ell},
\end{eqnarray*}
where
\begin{eqnarray*}
\alpha_\ell&=&\int\psi(x-\ell)\frac{\ud}{\ud x}\psi(x)\ud x\\
\beta_\ell&=&\int\psi(x-\ell)\frac{\ud}{\ud x}\varphi(x)\ud x\\
\gamma_\ell&=&\int\varphi(x-\ell)\frac{\ud}{\ud x}\psi(x)\ud x
\end{eqnarray*}
then by using refinement equations
we have in terms of filters
$(h_k,g_k)$:
\begin{eqnarray*}
\alpha_j&=&2\sum^{L-1}_{k=0}\sum^{L-1}_{k'=0}g_kg_{k'}r_{2i+k-k'},\\
\beta_j&=&2\sum^{L-1}_{k=0}\sum^{L-1}_{k'=0}g_kh_{k'}r_{2i+k-k'},\\
\gamma_i&=&2\sum^{L-1}_{k=0}\sum^{L-1}_{k'=0}h_kg_{k'}r_{2i+k-k'},
\end{eqnarray*}
where $r_\ell=\int\varphi(x-\ell)\frac{\ud}{\ud x}\varphi(x)\ud x, \ell\in Z.$
Therefore, the representation of $d/dx$ is completely determined by the
coefficients $r_\ell$ or by representation of $d/dx$ only on
the subspace $V_0$. The coefficients $r_\ell, \ell\in Z$ satisfy the
following system of linear algebraical equations
$$
r_\ell=2\left[ r_{2l}+\frac{1}{2} \sum^{L/2}_{k=1}a_{2k-1}
(r_{2\ell-2k+1}+r_{2\ell+2k-1}) \right]
$$
and $\sum_\ell\ell r_\ell=-1$, where $a_{2k-1}=$
$2\sum_{i=0}^{L-2k}h_i h_{i+2k-1}$, $k=1,\dots,L/2$
are the autocorrelation coefficients of the filter $H$.
If a number of vanishing moments $M\geq 2$ then this linear system of equations
has a unique solution with finite number of non-zero $r_\ell$,
$r_\ell\ne 0$ for $-L+2\leq\ell\leq L-2, r_\ell=-r_{-\ell}$.
For the representation of operator $d^n/dx^n$ we have the similar reduced
linear system of equations.
Then finally we have for action of operator $T_j(T_j:V_j\rightarrow V_j)$
on sufficiently smooth function $f$:
$$
(T_j f)(x)=\sum_{k\in Z}\left(2^{-j}\sum_{\ell}r_\ell f_{j,k-\ell}\right)
\varphi_{j,k}(x),
$$
where $\varphi_{j,k}(x)=2^{-j/2}\varphi(2^{-j}x-k)$ is wavelet basis and
$$
f_{j,k-1}=2^{-j/2}\int f(x)\varphi(2^{-j}x-k+\ell)\ud x
$$
are wavelet coefficients. So, we have simple linear para\-met\-rization of
matrix representation of our differential operator in wavelet basis
and of the action of
this operator on arbitrary vector in our functional space.
Then we may use such representation in all preceding sections.

We are very grateful to M.~Cornacchia (SLAC),
W.~Her\-r\-man\-nsfeldt (SLAC)
Mrs. J.~Kono (LBL) and
M.~Laraneta (UCLA) for
 their permanent encouragement.


\begin{thebibliography}{10}

\bibitem{1}
Fedorova, A.N., Zeitlin, M.G.
'Wavelets in Optimization and Approximations',
{\it Math. and Comp. in Simulation}, {\bf 46}, 527-534 (1998).
\bibitem{2}
Fedorova, A.N., Zeitlin, M.G.,
'Wavelet Approach to Polynomial Mechanical Problems',
New Applications of Nonlinear and Chaotic Dynamics in Mechanics,
Kluwer, 101-108, 1998.
\bibitem{3}
Fedorova, A.N., Zeitlin, M.G.,
'Wavelet Approach to Mechanical Problems. Symplectic Group,
Symplectic Topology and Symplectic Scales',
New Applications of Nonlinear and Chaotic Dynamics in Mechanics,
Kluwer, 31-40, 1998.
\bibitem{4}
Fedorova, A.N., Zeitlin, M.G
 'Nonlinear Dynamics of Accelerator via Wavelet
Approach', AIP Conf. Proc., vol.~{\bf 405}, 87-102, 1997,
Los Alamos preprint, phy\-sics/9710035.
\bibitem{5}
Fedorova, A.N., Zeitlin, M.G, Parsa, Z.,
'Wavelet Approach to Accelerator Problems', parts 1-3, Proc. PAC97,
vol.~{\bf 2}, 1502-1504, 1505-1507, 1508-1510, IEEE, 1998.
\bibitem{6}
Fedorova, A.N., Zeitlin, M.G, Parsa, Z.,
'Nonlinear Effects in Accelerator Physics: from Scale to Scale via Wavelets',
'Wavelet Approach to Hamiltonian, Chaotic and Quantum Calculations
in Accelerator Physics',
Proc. EPAC'98, 930-932, 933-935, Institute of Physics, 1998.
\bibitem{7}
Fedorova, A.N., Zeitlin, M.G., Parsa, Z.,
'Variational Approach in Wavelet Framework to Polynomial
Approximations of Nonlinear Accelerator Problems',
AIP Conf. Proc., vol.~{\bf 468}, 48-68, 1999.\\
Los Ala\-mos preprint, physics/9902062.
\bibitem{8}
Fedorova, A.N., Zeitlin, M.G., Parsa, Z.,
'Symmetry, Ha\-mil\-tonian Problems and Wavelets in
Accelerator Physics',
AIP Conf.Proc., vol.~{\bf 468}, 69-93, 1999.\\
Los Alamos preprint, physics/9902063.
\bibitem{9}
G.B.~Folland, 'Harmonic Analysis in Phase Space', Princeton, 1989.
\bibitem{10}
G.~Beylkin, R.R.~Coifman, V.~Rokhlin, Comm. Pure and Appl. Math., {\bf 44}, 141-183, 1991.


\end{thebibliography}
\end{document}